\documentclass[a4paper,onecolumn,11pt]{revtex4-2}
\pdfoutput=1

\usepackage[a4paper,top=2cm,bottom=2cm,left=2cm,right=2cm,marginparwidth=1.75cm]{geometry}

\usepackage{float}
\makeatletter
\let\newfloat\newfloat@ltx
\makeatother
\usepackage{natbib}

\usepackage{amsmath}
\usepackage{amssymb}
\usepackage{graphicx}
\usepackage[skip=12pt, indent=0pt]{parskip}
\usepackage{braket}
\usepackage{quantikz}
\usepackage{algorithm}
\usepackage{algpseudocode}
\usepackage{booktabs}
\usepackage[colorlinks=true, allcolors=blue]{hyperref}

\begin{document}

\title{Quantum optimisation applied to the Quadratic Assignment Problem}

\author{Andrew Freeland}
\author{Jingbo Wang}
\email[]{jingbo.wang@uwa.edu.au}

\affiliation{\large Centre for Quantum Information, Simulation and Algorithms, The University of Western Australia, Perth WA 6009, Australia}

\begin{abstract}

This paper investigates the performance of the emerging non-variational Quantum Walk-based Optimisation Algorithm (NV-QWOA) \cite{Bennett2024a, Bennett2024b, Bennett2025} for solving small instances of the Quadratic Assignment Problem (QAP). NV-QWOA is benchmarked against classical heuristics, namely the Max-Min Ant System (MMAS)~\cite{STUTZLE2000889} and Greedy Local Search (GLS)~\cite{christiansen2025instancespaceanalysisquadratic, Bennett2025}, as well as Grover’s quantum search algorithm~\cite{Grover1997}, which serves as a quantum baseline. Performance is evaluated using two metrics: the number of objective function evaluations and the number of algorithm iterations required to consistently reach optimal or near-optimal solutions across QAP instances with $n = 5$ to $n = 10$ facilities. The motivation for this study stems from limitations of both classical exact methods and current quantum algorithms. Variational Quantum Algorithms (VQAs), such as QAOA and VQE, while widely studied, suffer from costly parameter tuning and barren plateaus that hinder convergence. By adopting a non-variational approach, this work explores a potentially more efficient and scalable quantum strategy for combinatorial optimisation. The results provide a direct comparative analysis between classical and quantum frameworks, characterizing the average-case performance of NV-QWOA. Our findings highlight the practical utility of quantum walks for complex combinatorial problems and establish a foundation for future quantum optimisation algorithms.

\end{abstract}

\maketitle

\section{Introduction}

Most industries face difficult Combinatorial Optimisation Problems (COP), and developing strategies to solve them at scale remains a major challenge. Among the hardest of these COP problems are those classified as nondeterministic polynomial-time hard (NP-hard) in computational complexity theory, which are widely believed to be intractable for classical algorithms under the assumption that $P \neq NP$. This is because the solution space grows exponentially with problem size, and opportunities to exploit problem structure for acceleration are limited.

One COP of particular interest is the \textit{Quadratic Assignment Problem (QAP)}, which is considered as one of the most difficult NP-hard problems. Even finding a small $\epsilon$-approximation is NP-hard, as a polynomial-time approximation scheme would imply $P = NP$~\cite{sahni1976}. The QAP is versatile and can be adapted to both constrained and unconstrained settings. It is most commonly formulated as a facility-location assignment problem and has well-documented applications in hospital management~\cite{elshafei1977}, scheduling~\cite{geoffrion1976}, and backboard wiring~\cite{steinberg1961}. These applications are frequently cited in QAP research for their real-world relevance. The QAPLIB database~\cite{qaplib} provides a repository of benchmark instances that have guided classical algorithm development and remain a standard for performance evaluation.

This research benchmarks the performance of the emerging non-variational \textit{Quantum Walk-based Optimisation Algorithm (NV-QWOA)} in solving QAP instances. We compare NV-QWOA against established classical heuristics, namely the Max-Min Ant System (MMAS)~\cite{STUTZLE2000889} and Greedy Local Search (GLS)~\cite{christiansen2025instancespaceanalysisquadratic, Bennett2025}, as well as Grover’s quantum search algorithm~\cite{grover1996}, which serves as a quantum baseline. The primary metric for comparison with classical heuristics is the number of objective function evaluations, while measurement and solution probabilities are also considered for NV-QWOA versus GLS. For Grover’s algorithm, we compare the number of quantum iterations required as a function of problem size. 



\subsection{Classical algorithms}

Classical algorithms for solving QAPs and other combinatorial optimisation problems (COPs) generally fall into two categories: exact algorithms and heuristics. Exact algorithms explore the solution space exhaustively, systematically eliminating low-quality solutions to guarantee finding the global optimum. In contrast, heuristics search the solution space until no further improvement is possible. While heuristics do not guarantee optimality and often converge to local optima, they are significantly more efficient than exact algorithms. Further details on exact algorithms and heuristics, particularly the Max-Min Ant System (MMAS) and Greedy Local Search (GLS) for QAP, are provided in Section~\ref{background-methods}.

Shortly after the foundational paper defining the QAP \cite{kooopmans1957}, Gilmore and Lawler \cite{gilmore1962, lawler1963} independently observed that a lower bound on the QAP can be obtained by reducing it to an $n^2 \times n^2$ Linear Assignment Problem (LAP), with the additional constraint that the LAP permutation matrix $L$ must be the Kronecker square of the QAP permutation matrix $X$: $L = X^{\otimes 2}$. This lower bound, known as the Gilmore–Lawler bound (GLB), has time complexity $O(n^3)$ but deteriorates in performance as $n$ increases. GLB has inspired numerous relaxations and reductions that accelerate branch-and-bound algorithms \cite{gavett1966}, cutting-plane methods \cite{bazaraa1980}, and polyhedral approaches \cite{padberg1991}. More advanced bounds, such as the semi-definite relaxation–matrix splitting (SDRMS) bound \cite{peng2010}, use relaxations and reformulations to achieve smaller optimality gaps for large and difficult instances, including Taillard’s benchmarks for $n > 50$ \cite{taillard1991, taillard1995, qaplib}. While these bounds outperform GLB in terms of gap size, they are typically more computationally expensive, creating a trade-off between runtime and solution quality.

The best-known exact algorithms for QAP are branch-and-bound (B\&B) methods. These algorithms can solve instances up to about $n = 30$, but require substantial parallel computation time \cite{anstreicher2002}. Exact methods are most suitable for small to moderate problem sizes, where they guarantee optimality with competitive runtimes. They also play a critical role in generating optimal solutions for benchmark datasets such as QAPLIB.
Heuristics are generally divided into local-search and population-based categories. Examples include greedy local search (GLS) \cite{merz2002}, iterated local search (ILS) \cite{ils}, simulated annealing \cite{kirkpatrick1983}, tabu search \cite{GLOVER1986533}, ant colony optimisation (ACO) \cite{dorigo1992}, and evolutionary or genetic algorithms \cite{bartzbeielstein2014, moscato2000}. Population-based algorithms often incorporate local-search subroutines to quickly find local minima and typically return near-optimal solutions with small gaps, though they are more computationally intensive \cite{STUTZLE2000889}.

\subsection{Quantum algorithms}

Quantum combinatorial optimisation is an emerging area of research motivated by the need to solve hard problems such as the Quadratic Assignment Problem (QAP). Early approaches employed the Quantum Adiabatic Algorithm (QAA)~\cite{farhi2000, farhi2001}, which gradually evolves an initial Hamiltonian into a final Hamiltonian whose ground state encodes the optimal solution. Under sufficiently slow evolution, the system remains in its ground state, ensuring that measurement yields the optimal solution with high probability.
Farhi et al.~\cite{farhi2014quantumapproximateoptimizationalgorithm} later introduced the Quantum Approximate Optimisation Algorithm (QAOA), a gate-based method that contrasts with the adiabatic approach. QAOA uses a parameterized quantum circuit alternating between two Hamiltonians, with $2p$ parameters optimized through a classical feedback loop. Its performance depends heavily on parameter initialisation and circuit depth, as deeper circuits improve expressibility but increase computational cost~\cite{bittel2021}. QAOA is also susceptible to barren plateaus—regions of the cost landscape with vanishing gradients that hinder convergence~\cite{Larocca2024ReviewBP}.

Marsh and Wang~\cite{Marsh2019, Marsh2020, marsh2021} generalized QAOA into the \textit{Quantum Walk-based Optimisation Algorithm (QWOA)}. QWOA replaces the transverse-field Hamiltonian with a continuous-time quantum walk (CTQW)~\cite{QWbook2014, LokeWang2017}, which connects only feasible solutions within the search space. This design prevents amplitude flow to invalid states, improving constraint handling compared to penalty-based methods~\cite{Bennett2021quantum, slate2021quantum, matwiejew_quantum_2023, Qu2024}.
Building on QWOA, Bennett et al.~\cite{Bennett2024a, Bennett2024b, Bennett2025} proposed the \textit{Non-Variational Quantum Walk-based Optimisation Algorithm (NV-QWOA)}, which integrates CTQW-based mixers with QAOA principles to exploit problem structure.  Unlike conventional QAOA/QWOA algorithms requiring extensive parameter tuning, NV-QWOA requires only three hyperparameters governing evolution time and walk dynamics. This reduction eliminates costly classical optimisation loops, enhancing scalability and mitigating barren plateau phenomena~\cite{Bennett2024a, Bennett2024b, Bennett2025}. Consequently, NV-QWOA represents a promising direction for resource-efficient quantum optimisation with reduced susceptibility to training pathologies.

\section{Theoretical background} \label{background-methods}

This section details the mathematical description of QAPs and the algorithmic processes behind the classical and quantum heuristics considered for comparison. The classical algorithms considered are the Max-Min Ant System and Greedy Local Search. The quantum algorithms considered are the NV-QWOA and Grover's quantum search algorithm, of which both offer a more "non-variational" approach to solving COPs compared to the large-scale parameter tuning and high computational expenses of VQAs such as QAOA, VQE and the QWOA. Background into the QAOA and QWOA are also given due to their relevance in the development of the NV-QWOA.

\subsection{Quadratic Assignment Problem}
Quadratic Assignment Problems (QAPs) were first defined in 1957 by Koopmans and Beckmann as a mathematical model related to indivisible economic activities, assigning production plants to locations in context of maximising revenue \cite{kooopmans1957}. The computational difficulty of the problem was quickly recognized, and methods were devised to establish lower bounds on the cost landscape and to inform classical algorithm design \cite{gavett1966, bazaraa1980}.

\begin{figure}
    \centering
    \includegraphics[width=0.4\linewidth]{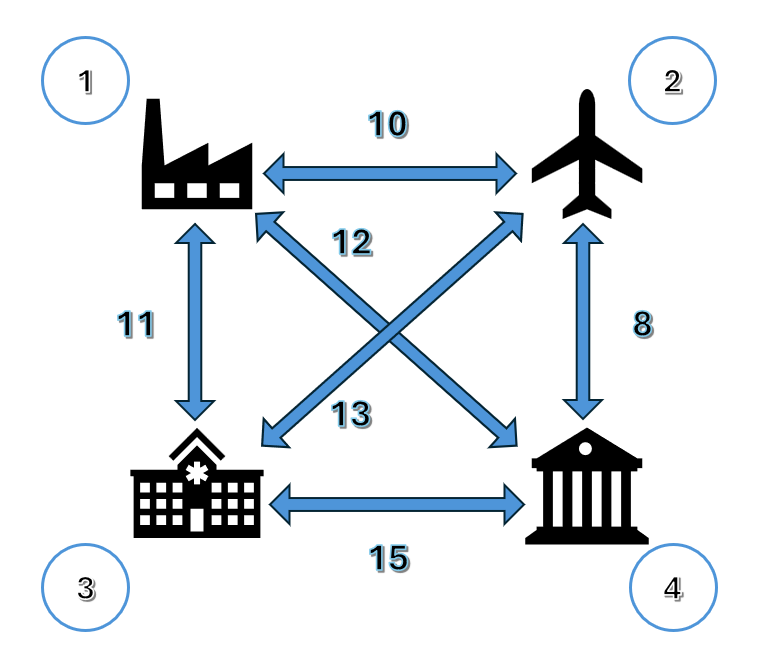}
    \includegraphics[width=0.4\linewidth]{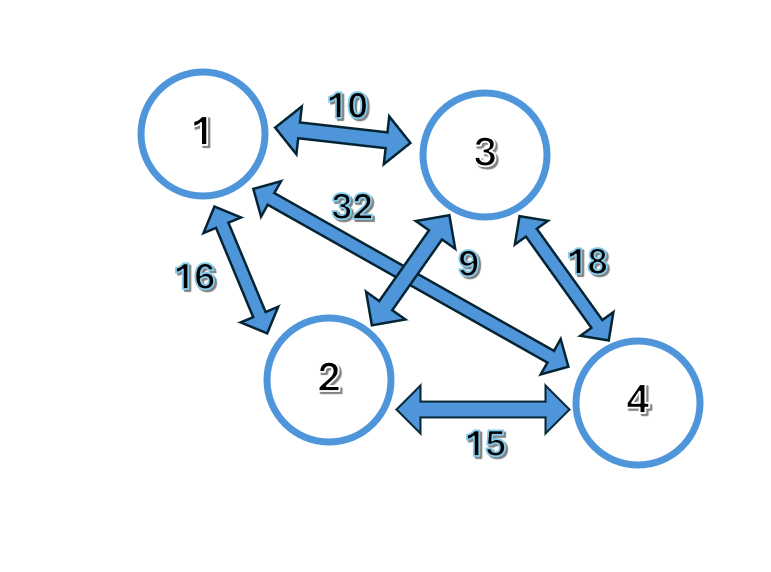}
    \caption{\textit{Facility flows and location distances of an example $n = 4$ QAP}}
    \label{fig:example-qap}
\end{figure}

The QAP is defined as the problem of assigning a set of facilities to a set of locations so as to minimize the total cost of assignment. The total assignment cost is given by the sum of the products of flows between facilities $f_{i,j}$ with distances between their assigned locations $d_{\pi(i),\pi(j)}$. An example QAP can be seen in Figure \ref{fig:example-qap}, along with the respective flow and distance matrices $F$ and $D$ in Equations \ref{eq:flow-matrix} and \ref{eq:dist-matrix} below: 
\begin{equation} \label{eq:flow-matrix} 
    F = 
    \begin{bmatrix}
        0 & 10 & 11 & 12 \\
        10 & 0 & 13 & 8 \\
        11 & 13 & 0 & 15 \\
        12 & 8 & 15 & 0
    \end{bmatrix},
\end{equation}
\begin{equation} \label{eq:dist-matrix} 
    D = 
    \begin{bmatrix}
        0 & 16 & 10 & 32 \\
        16 & 0 & 9 & 15 \\
        10 & 9 & 0 & 18 \\
        32 & 15 & 18 & 0
    \end{bmatrix}.
\end{equation}

The principle behind minimizing the total assignment cost is that facilities with large flows between them should be assigned to locations that are close to each other, while facilities with minimal flow can be placed farther apart. Both the flow and distance matrices are typically symmetric and have zero diagonals; however, asymmetric instances have also been studied \cite{qaplib, loiola2007}. In this paper, only symmetric matrices are considered.

The objective function of the symmetric QAP is given below, with the aim of minimisation:
\begin{equation} \label{eq:qap-objective}
    f(x) = \sum_{i=1}^{n}{\sum_{j=1}^{n}{f_{i,j}d_{\pi(i),\pi(j)}}},
\end{equation} 
where $f_{i,j}$ is the flow between facility $i$, and $j$, $d_{\pi(i), \pi(j)}$ is the distance between locations $\pi(i)$ and $\pi(j)$, and $\pi: \{1, 2, ..., n\} \rightarrow \{1, 2, ..., n\}$ represents the permutation that maps the facilities to locations. 
The number of possible solutions to a given QAP instance cooresponds to the number of permutations of the assignments of facilities to locations in the set of all possible permutations $S_n = \{\pi_i \}_{i=1}^{N}$. Given $n$ facilities and locations, which define the problem size of the QAP as $n$, there are $N = n!$ possible solutions, with the factorial growth of number of solutions contributing to the $NP$-hardness of the QAP. 

In order to be able to extract solutions given costs, the permutations need to be appropriately indexed, since the factorial growth of the solution space prevents ease of tracking and storing solutions, and since indexes will map solutions to costs uniquely. The indexing algorithm used for the paper is outlined in Algorithm \ref{alg:perm-index}, and the un-indexing algorithm that returns the permutation corresponding to an obtained index, useful for finding the solution that corresponds to the optimal cost, is outlined in Algorithm \ref{alg:perm-unindex}.

\begin{algorithm}
\caption{Permutation indexing}\label{alg:perm-index}
\begin{algorithmic}[1]
\Require Problem size $n$, permutation $\pi_i \in S_n$
\Ensure Index $i$ of $\pi_i$
\State Initialise $i \leftarrow 0$
\State Generate list of elements $\phi \leftarrow \{0,1,...,n-1\}$
\State Generate $factorials = \{0!, 1!, ..., (n-1)!\}$
\State Initialise $depth \leftarrow n-1$
\For{each $k\in \pi_i$}
\State Obtain index $l$ of $k$ in $\phi$
\State Remove $k$ from $\phi$
\State $i \leftarrow i + l \cdot factorials \lbrack depth \rbrack$
\State $depth \leftarrow depth - 1$
\EndFor
\end{algorithmic}
\end{algorithm}

\begin{algorithm}
\caption{Permutation un-indexing}\label{alg:perm-unindex}
\begin{algorithmic}[1]
\Require Problem size $n$, index $i$
\Ensure Permutation $\pi_i$
\State Initialise $\pi_i = \lbrack \rbrack$
\State Generate list of elements $\phi \leftarrow \{0,1,...,n-1\}$
\State Generate $factorials = \{0!, 1!, ..., (n-1)!\}$
\State Initialise $depth \leftarrow n-1$
\While{$\phi$ != $\lbrack \rbrack$}
\State $block \leftarrow factorials[depth]$
\State Pop $(i // block)$-th element in $\phi$ and append to $\pi_i$  
\State $i \leftarrow i \% block$
\State $depth \leftarrow depth - 1$
\EndWhile
\end{algorithmic}
\end{algorithm}

\subsection{Max-Min Ant System}

The Max-Min Ant System (MMAS) is a variant of the population-based ant colony optimisation (ACO) heuristic \cite{STUTZLE2000889}. In ACO algorithms, many agents known as ants are generated to construct new solutions to a given QAP instance by adding solution components (local facility-location assignments) from an initially empty solution, and depositing pheromones, encoded in a $n$-by-$n$ matrix $\tau_{i,j}$, to inform the next generation of ants of optimal and non-optimal solution regions, with more optimal solution components receiving a higher pheromone amount, and thus encouraging the next population of ants to use these solution components to try and find the optimal solution. Within MMAS, only a single ant is chosen to deposit new pheromones at each iteration, which may or may not correspond with the optimal solution. To avoid stagnating the search, all pheromone trails are decreased by a factor $\rho$ referred to as the pheromone evaporation rate. The ants are also guided by problem-dependent heuristic information, the indicator of the expected proficiency of a particular pairing of facility and location assignments, encoded in a $n$-by-$n$ matrix $\beta_{i,j}$. For the QAP, this may be based on effective lower bounds on completing the partial solution, usually the GLB \cite{maniezzo1994}, but it was found in \cite{STUTZLE2000889} that using heuristic information is not necessary due to its negligible effect in solution degradation. 

The MMAS parameters include: the number of ants $m$, pheromone importance $\alpha$, pheromone matrix $\tau_{i,j}$, initial pheromone level $\tau_0$, pheromone evaporation rate $\rho$, pheromone update constant $Q$, heuristic importance $\beta$, heuristic information matrix $\beta_{i,j}$, and the maximum number of iterations $T$.
The general schematic of the MMAS is given in Algorithm \ref{alg:ant-cap} as derived from \cite{STUTZLE2000889}:

\begin{algorithm}
\caption{MMAS Algorithm}\label{alg:ant-cap}
\begin{algorithmic}[1]
\Require Problem size $n$, flow and distance matrices $F$ and $D$, MMAS parameters
\Ensure Optimised solution cost $x$
\State $I \leftarrow 1$
\While{$I < T$}
\For{each ant $k$}
\State Build complete assignment $\pi_k$ step-wise through $p_{i,j}^k = \frac{(\tau_{i,j})^{\alpha}(\beta_{i,j})^{\beta}}{\sum_{l \in Available}(\tau_{i,l})^{\alpha}(\beta_{i,l})^{\beta}}$
\State $x \leftarrow f(\pi_k)$
\State $\tau_{i,j} \leftarrow (1-\rho)\tau_{i,j}$
\State $\tau_{i,j} \leftarrow \tau_{i,j} + \Delta \tau_{i,j}^{best},$ where $\Delta \tau_{i,j}^{best} =  
    \begin{cases}
    \frac{Q}{f(\pi_{best})} & \text{if $\pi_{best}(i)=j$} \\
    0 & \text{otherwise}
    \end{cases}$
\State $\tau_{i,j} \leftarrow \min(\tau_{max}, \max(\tau_{min}, \tau_{i,j}))$
\EndFor
\State $I \leftarrow I + 1$
\EndWhile
\end{algorithmic}
\end{algorithm}

The MMAS algorithm designed for the QAP is an extension on the algorithm designed for the TSP, simply modifying the solution construction steps to correlate with the facility-assignment formulation rather than a tour, as well as a lack of use of heuristic information $\beta_{i,j}$ due to its negligible effect in solution degradation. In this paper, we set the heuristic importance $\beta \leftarrow 0$.

\subsection{Greedy Local Search}

Greedy Local Search is a variant of a local search algorithm that starts with an random initial solution, and performs all possible transposition swaps between elements of the permutation to find the best-performing swap. Following the best swap, the best-swap finding procedure continues until no improving swap is found. This termination condition ensures the GLS finds a minimum, which is often a local minimum, since the number of local minima can scale dramatically with problem size $n$, as seen with the MaxCut problem in \cite{Bennett2025}. GLS is efficient in finding at least a local minimum, and performs $O(n^2)$ every iteration, with the number of objective functions evaluations $N_{f(x)} = iterations \cdot \frac{n(n-1)}{2}$, where the second term is the number of possible transpositions from a starting permutation. 

The GLS algorithm used is given in Algorithm \ref{alg:greedy-local-search}. The particular GLS used for comparison is not augmented by randomisation to escape local optima and investigate new landscape regions, but kept as a purely hill-climbing local search greedy algorithm to emphasise a contrasting point with the NV-QWOA, which exploits the problem structure for the problem of consideration to amplify optimal and near-optimal solutions precisely.

\begin{algorithm}
\caption{Greedy Local Search}\label{alg:greedy-local-search}
\begin{algorithmic}[1]
\Require Problem size $n$, flow and distance matrices $F$ and $D$ 
\Ensure Optimised solution cost $x$
\State Generate random permutation $\pi$
\State $x \leftarrow f(\pi)$
\State $improved \leftarrow True$
\While{$improved = True$}
\State $\Delta_{best} = 0$
\State $\Delta_{best}$ indices $\Delta_i = -1$ and $\Delta_j = -1$
\State $\Delta_{best} \leftarrow \Delta_{swap}(\pi,i,j)$ for $i = 0,1,...,n-2$ and $j = i+1,i+2,...,n-1$
\State Update $\Delta_i$ and $\Delta_j$
\If{$\Delta_{best} < 0$}
\State $\pi[\Delta_i, \Delta_j] \leftarrow  \pi[\Delta_j, \Delta_i] $
\State $x \leftarrow x + \Delta_{best}$
\EndIf
\If{$\Delta_{best} \geq 0$}
\State $improved \leftarrow False$
\EndIf
\EndWhile
\end{algorithmic}
\end{algorithm} 

\subsection{Quantum Approximate Optimisation Algorithm}
The Quantum Approximate Optimisation Algorithm (QAOA) \cite{farhi2014quantumapproximateoptimizationalgorithm} was initially developed to solve MaxCut problems, another COP of significant interest in quantum combinatorial optimisation. The cost function of a given problem $C(x)$ is encoded as a problem Hamiltonian $H_p = C$, whose ground-state corresponds to the optimal solution. Solutions $x$ are expressed as $k$-bit binary strings $b_0b_1...b_{k-1}$, and the algorithm prepares a parameterised quantum state to be optimised through minimisation or maximisation of the expectation value $\braket{\beta,\gamma|C|\beta,\gamma}$ , where $\beta = [\beta_1, \beta_2, ...\beta_p]$ and $\beta = [\gamma_1, \gamma_2, ...\gamma_p]$ are vectors of $p$ parameter angles.

The Quantum Alternating Operator Ansatz (QAOAz) framework \cite{Hadfield_2019} generalises the family of operations used in the QAOA, and involves mapping the ground-state of an initial Hamiltonian, usually the equal superposition state, to the ground-state of the given problem Hamiltonian. This mapping is achieved by sequentially applying a problem unitary composed of $H_p$ and $\gamma$ to affect cost quality, and a mixer unitary composed of a mixer Hamiltonian $H_m = \sum_{j=1}^{N} \sigma_{j}^{x}$ involving the promotion of the solution binary strings to quantum spins, resulting in the parameterised quantum state:
\begin{equation}
    \ket{\beta,\gamma} = \prod_{l=1}^{p} e^{-i\beta_{l}H_m} e^{-i\gamma_{l}H_c} \ket{s}^{\otimes k},
\end{equation}
where $\ket{s}$ is the superposition state over $k$ qubits to encode the solutions.  

\subsection{Quantum Walk-based Optimisation Algorithm}

The Quantum Walk-based Optimisation Algorithm (QWOA) \cite{Marsh2020} emerged from development of the QAOA and QAOAz, but instead defines the mixer unitary (Equation \ref{eq:mixer-eq}) as a continuous-time quantum walk (CTQW) over a mixing graph, and the problem unitary as a phase-shifting unitary (Equation \ref{eq:phase-shift-eq}) that separates solution states by phase proportional to the objective function value (also referred to as the quality or cost) of the solution states. These unitaries, like with the problem and mixing Hamiltonians in the QAOA, are sequentially applied to an initial equal superposition state that encodes all the solutions to the problem $p$ times. The mixing graph is initially a hypercube, with variable connectivity allowing to hone in on optimal solutions.

The QWOA also requires efficient indexing and unindexing algorithms to map solutions to mixing graph nodes, and thus perform the quantum operations to amplify the state for increased odds at measuring the optimal solution.
Following $p$ alternating unitary applications, the amplified state is given by:
\begin{equation}
    \ket{\gamma,t} = U_M(t_p)U_Q(\gamma_p)...U_M(t_2)U_Q(\gamma_2)U_M(t_1)U_Q(\gamma_1)\ket{s}.
\end{equation}

\subsection{Non-variational QWOA}

As a successor to the QAOA and QWOA, the NV-QWOA follows the procedure of alternately applying two unitaries $p$ times to drive an initial equal superposition state of all solution states to a final state with significant amplification applied to the optimal solution and near-optimal solutions, and significant attenuation applied to low-quality solutions \cite{Bennett2025, Bennett2024a, Bennett2024b}. Then ultimately, a small number of repeated state preparations and measurements within a shot budget can be made to measure the optimal solution. The difference of the NV-QWOA with the QWOA and other VQAs lies in the ability to select the relevant mixer (mixing graph) for the problem of interest, and how the phases of solutions states combine to constructively interfere for near-optimal solutions, and destructively interfere for suboptimal ones, with only optimisation of 3 parameters as opposed to 2$p$. 

Within this framework, QAP solutions are embedded within the computational basis states of an input register through a one-hot encoding scheme, by assigning a sub-register to each of the $n$ variables $x_j$, and identifying the $n$ computational basis states of each of these sub-registers which directly encode the variable values $x_j \in \{0, 1, ..., n - 1\}$, thus requiring $n^2$ total qubits. The states of the sub-registers depend on each other due to the non-repetition constraint of permutations (all elements are unique), with a equal superposition preparation scheme of complexity $O(n^3)$ developed in \cite{Bennett2024b}. The initial quantum state encoding all solution states of the form $\ket{\mathbf{x}} = \prod_{j=0}^{N-1}{\ket{x_j}}$ in equal superposition is given by:
\begin{equation}\label{eq:initial-sup-eq}
    \ket{s} = \frac{1}{\sqrt{N}} \sum_{\mathbf{x} \in S_n}{\ket{\mathbf{x}}}
\end{equation}

The NV-QWOA features $p$ layers of alternating unitary applications of the phase-shifting unitary $U_Q$ and mixing unitary $U_M$. For each layer, $U_Q$ introduces a phase $e^{-i\gamma Q}$ to each solution state proportional to its objective function value contained within the quality matrix $Q$ and parameter $\gamma$, expressed below:
\begin{equation}\label{eq:phase-shift-eq}
    U_Q(\gamma) = e^{-i\gamma Q}.
\end{equation}

The mixing unitary then disperses amplitude probabilities between neighbouring solutions for some walk time $t$ as below:
\begin{equation}\label{eq:mixer-eq}
    U_M(t) = e^{-itA},
\end{equation}
where $A$ is the adjacency matrix defining the connectivity of solution states as vertices of the graph. Transposition mixing graphs define nearest-neighbours as solutions that differ by one transposition swap (swapping 2 permutation elements), equivalent to the solutions being a Hamming distance of 2 away. $A$ is then built from iterative construction of neighbours from an initial vertex/permutation. 

The overall action of $U_M(t)U_Q(\gamma)$ across the full circuit depth $p$ is expressed in Equation \ref{eq:NV-QWOA-symb} below:
\begin{equation}\label{eq:NV-QWOA-symb}
    \ket{\gamma, t, \beta} = \left[\prod_{i = 0}^{p - 1} U_{M}(t_{i}) U_{Q}\left(\frac{-\gamma_{i}}{\sigma}\right)\right] \ket{s}
\end{equation}

Each $\gamma_i$ and $t_i$ value is defined by a linear ramping independent of $p$ and dependent on $\beta$, with $\gamma_i$ linearly increasing over domain $[\beta \gamma,\gamma]$, and $t_i$ decreasing over domain $[\beta t,t]$:
\begin{equation}\label{eq:linear-ramping}
\begin{split}
    & \gamma_i = \left(\beta + (1 - \beta)\frac{i}{p-1}\right) \gamma, \\
    & t_i = \left(1 - (1 - \beta)\frac{i}{p-1}\right) t .
\end{split}
\end{equation}

Based on statistical inference, each raw $\gamma$ value is scaled by the standard deviation derived from sample experiments \cite{Bennett2024b}. 

The full algorithm of the NV-QWOA, as would be used for quantum hardware, is given as follows: 

\begin{algorithm}
\caption{Non-variational QWOA procedure} \label{alg:NV-QWOA}
\begin{algorithmic}[1]
\State Initialise equal superposition of feasible solutions $\ket{s}$, with each $\ket{x}$ allocated a sub-register of $n$ qubits as in \ref{eq:initial-sup-eq}
\State Repeatedly apply \ref{eq:phase-shift-eq} and \ref{eq:mixer-eq} alternately $p$ times with $\{\gamma_i,t_i\}$ to obtain amplified state $\ket{\psi}$ from \ref{eq:NV-QWOA-symb}
\State Measure $\ket{\psi}$ to return approximate solution $x^*$, of which its cost/quality can be quickly determined by $f(x)$
\State Repeat steps 1-3 for number of desired shots (state preparations and measurements) to approximate $\braket{\gamma,t,\beta|Q|\gamma,t,\beta}$
\State Optimal solution is defined as solution corresponding to minimum cost/expectation value
encountered, ideally the global minimum
\end{algorithmic}
\end{algorithm}

\begin{figure}[htb]
    \centering
    \includegraphics[width=0.5\linewidth]{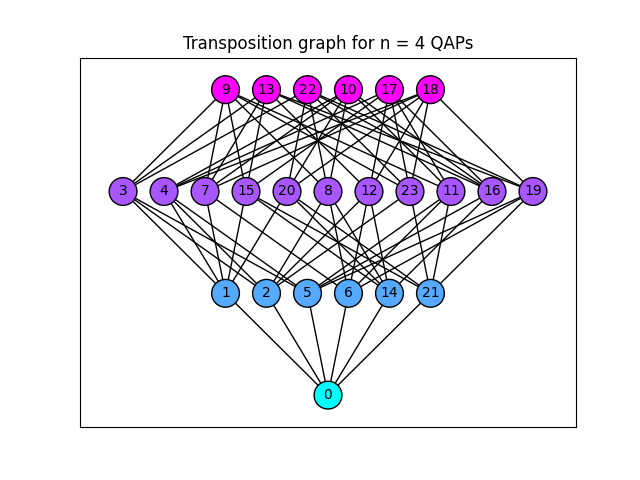}
    \caption{\textit{The transposition graph for $n = 4$ QAP instances, centered at solution $0 = [0,1,2,3]$. The nodes are the solutions, and the edges connect nearest-neighbour solutions i.e those that differ by one transposition.}}
    \label{fig:ex-transposition}
\end{figure}

The appropriate choice of mixing graph for the QAP is the transposition graph, since nearest neighbour solutions are transpositions of each other; an example transposition mixing graph for $n = 4$ QAP instances is given in Figure \ref{fig:ex-transposition}. The mixing graph is constructed through the adjacency matrix 
\begin{equation}
    A = \sum_{i=0}^{n-2} \sum_{j=i+1}^{n-1} SWAP_{i,j},
\end{equation}
where the $SWAP$ operator is a permutation matrix that swaps the states of the $i^{th}$ and $j^{th}$ registers and applies the identity operation to the remaining registers. $A$ itself is not unitary, but the mixing operation itself $U_M(t) = e^{-itA}$ is unitary.

\section{Methodology}

This section describes the detailed approach used to perform NV-QWOA simulations, generate benchmark metrics for all heuristics, and create QAP instances. Small problem sizes ($n = 5, 6, 7, 8, 9, 10$) were selected due to parallel computation constraints when simulating NV-QWOA runs for benchmarking and comparison. Details on simulation complexity and runtime are provided in Section~\ref{results}.
Classical heuristics were chosen for comparison rather than exact algorithms, even though exact methods can be applied to small problem sizes and guarantee optimal solutions. This choice ensures benchmarking against algorithms that share a similar non-exhaustive approach. Unlike exact algorithms, heuristics do not systematically explore the entire solution space or eliminate candidates until the global optimum is found. Instead, they typically converge to local minima, which may correspond to suboptimal solutions.
The primary metrics for comparing NV-QWOA with classical heuristics are the number of objective function evaluations and measurement/solve probabilities. For classical heuristics, the number of evaluations correlates directly with execution time, as more evaluations require more processing. Additional NV-QWOA benchmarks include analyzing how the required circuit depth $p$ scales with $n$ to achieve a mean 10\% optimal solution probability (OSP) compared to Grover’s search, as well as the expected mean internode distance across problem sizes.

\subsection{MMAS configuration}

For the MMAS algorithm, it is inherently designed to heuristically solve moderately sized instances more efficiently than the state-of-the-art exact algorithms as well as problem sizes much larger than the limit achieved by exact algorithms. As such, MMAS application for smaller problem sizes is known to be efficient. Additionally, the usual termination conditions of the MMAS algorithm involve reaching a certain number of iterations of the main execution loop or time-limit. Since the time-limit is correlated with number of iterations and objective function evaluations, the number of iterations to terminate the search is of consideration, and since the MMAS is more efficient at finding instances for small problem sizes considered here, it was decided that the algorithm would execute until the optimal solution is found for each instance. 

The number of objective function evaluations without factoring in local search is given by the product of the number of MMAS iterations required to find the optimal solution with the number of ants used in the algorithm: $N_{f(x)} = m*I$. To develop a greater statistical picture on the distribution of $N_{f(x)}$ across problem sizes, 50 algorithm runs were executed to derive the mean $N_{f(x)}$ for each of the 30 QAP instances, with results seen in Figure \ref{fig:mmas-obj-func-evals}.

When computing the objective function evaluations for the MMAS algorithm, each of the 50 runs per problem instance were initially computed sequentially and lead to longer than necessary computation times, before OpenMPI was employed for nodes dedicated to a smaller subset of runs to run in parallel, significantly improving computation time. 

\subsection{NV-QWOA configuration}

With the aim to benchmark the performance of the NV-QWOA in solving QAPs compared to the classical heuristics, it is important to understand how the number of ansatz layers or circuit depth $p$ changes with the problem size $n$. For efficient hardware implementations, $p$ should be polynomially bounded in $n$. To gauge an optimal $p$ value to appropriately amplify the probability of measuring the optimal solution, a 10\% threshold is used, since the probability of failure to measure the optimal solution after 1000 circuit shots is of the order $10^{-46}$. The 1000 independent state preparations and measurements of the amplified quantum state following execution of the algorithm $p$ times is within the typical range of quantum circuit evaluations for NISQ-era hardware \cite{Bennett2025}. The optimal $p$ value then informs the number of objective function evaluations as $s*p$ which can be directly compared against the classical heuristics. In line with the benchmarking of the NV-QWOA for solving the MaxCut problem \cite{Bennett2025}, we only consider 4 shots, thus defining the number of objective function evaluations performed by the NV-QWOA as $4p$. 

The 2$p$ variational parameters $\{\gamma_1, \gamma_2, ..., \gamma_p, t_1, t_2, ..., t_p \}$ that define VQAs are substituted with only 3 tunable parameters $\{\gamma, t, \beta \}$ for the NV-QWOA. These define the initial parameters that act with the first layer of alternating unitary application $U_M(t)U_Q(\gamma)$, and construct the full set of parameters without fine-tuning through linear ramping seen in Equation \ref{eq:linear-ramping}. These parameter values are tracked at each layer, and the evolution for increasing $p$ can be observed for each problem instance, with results tabulated in Table \ref{tab:sim-full-results}.

\subsection{Solution shell structure}

For a given problem size $n$, the solution shell $k$ contains all solutions at permutation distance $k$ away from an arbitrarily chosen reference solution consists of permutations that contain $n-k$ disjoint cycles. These disjoint cycles represent chains of element mappings between a reference permutation and the permutation of interest, terminating when the mapping returns to the starting element. An $l$-cycle, therefore, is defined as a cycle that requires $l$ steps to return to its initial element under the permutation mapping.

For instance, considering the permutation $2310$ from fixed vertex $0123$, the disjoint cycle is a 4-cycle. We start at element $0$ to perform the mappings $\pi(0) \rightarrow 3, \pi(3) \rightarrow 1, \pi(1) \rightarrow 2, \pi(2) \rightarrow 0$. This process continues with elements $1, 2, 3$ to determine the number of disjoint cycles present for that permutation, and thus determining its solution shell location. Solutions that are one transposition away from the fixed vertex have $n-1$ disjoint cycles. Additionally, subshells are defined not only by number of disjoint cycles, but also the type of cycles present. For this paper however, only regular shells were considered. 

A point of interest is to see how the probabilities of solution shells, with respect to distance away from the optimal solution, and defined by the transposition graph connectivity, change before and after execution of the NV-QWOA, to illustrate a broader picture of how optimal and near-optimal solutions are amplified across problem size, and how poor solutions are attenuated. 


All 30 QAP instances were randomly generated using uniform distributions in Python's NumPy package. 2D location grids were created before initialising the distance matrices based on the Euclidean distance between points. Flow matrix entries were multiplied by a factor of 20 after random initialisation. Excessive phase rotations for constructive and destructive interference in $U_Q$ are avoided by scaling the encoded costs within the $(0, 2\pi)$ region, the natural range for quantum phase angles.
The instance generation was performed on Pawsey Supercomputing Research Centre’s Setonix Supercomputer system, utilising OpenMPI parallelisation with a small number of tasks for the larger problem sizes (namely $n = 10$), and were stored as .npy files. 


\section{Results and discussion} \label{results}

This section examines the results obtained from performing NV-QWOA simulations and applying classical heuristics to solve Quadratic Assignment Problems (QAPs). An important consideration when comparing heuristics is that finding the global optimal solution is not guaranteed, unlike with exact algorithms. Therefore, average-case and worst-case performance metrics define the quality of these heuristics when solving unseen QAP instances for real-world applications, particularly when aiming to obtain optimal or near-optimal solutions with a small optimality gap efficiently. Simulation results are detailed before presenting the main benchmarking results for classical and quantum heuristics.

The optimiser chosen to optimise the parameter vector was the Nelder-Mead algorithm within SciPy's minimize function \cite{scipy}. Initially, the COBYLA algorithm was used, but resulted in much longer convergence times for minimal, or otherwise non-existent, improvements to the objective function expected value minimisation. This Nelder-Mead optimiser recorded the number of objective function evaluations for each layer up to the circuit depth $p$.

Parallel computation jobs were submitted through the SLURM scheduler \cite{slurm-cite}. The \textit{--array} directive was utilised to perform simulations for each instance across problem sizes, and results were generated up to a suitable $p$ to ensure the NV-QWOA returned a mean 10\% OSP across the instances for each problem size. Results for each problem size were generated sequentially, and all optimiser results (including $\{\gamma, t, \beta\}$ parameter values and optimiser objective function evaluations), number of iterations $p$, objective function values, and simulation times, were all stored in CSV files for each problem instance across the problem sizes.

\subsection{Mean Simulation Time} \label{sim-results}
The mean simulation time required to execute NV-QWOA for 30 randomly generated QAP instances across problem sizes $n \in \{5, 6, 7, 8, 9, 10\}$, along with the optimized parameter values $\{\gamma, t, \beta\}$ and their corresponding standard deviations, is summarized in Table~\ref{tab:sim-full-results}. The reported standard deviations quantify the variability in both simulation times and parameter estimates.

As shown in Table~\ref{tab:sim-full-results}, simulation time exhibits approximately linear growth for smaller instances but escalates sharply as $n$ increases, despite an increased allocation of MPI tasks. This behavior suggests that the computational cost of classical simulation scales exponentially with problem size, underscoring the inherent limitations of simulating quantum algorithms on classical hardware. Furthermore, the widening range of simulation times for larger $n$, as reflected by the standard deviations, reinforces this observation and highlights the growing impact of resource constraints on scalability.

\begin{table}[ht]
    \centering
    \begin{tabular}{cllccllllll}\toprule
          & &&  \multicolumn{2}{c}{Simulation time (s)} & \multicolumn{6}{c}{Parameter values}\\\midrule
         $n$ & $p$&MPI &  $\bar{t_s}$& $\sigma_{t_s}$& $\bar{\gamma}$& $\sigma_{\gamma}$& $\bar{\beta}$& $\sigma_{\beta}$& $\bar{t}$&$\sigma_t$ \\
         5 & 3&16&  1.23&  0.10& 1.5197& 0.9714& 0.4533& 0.0990& 0.3153&0.0423\\
         6 & 6&16&  4.05&  0.29& 1.8081& 1.9879& 0.3623& 0.0808& 0.2155&0.0258\\
         7 & 10&16&  15.52&  0.61& 1.5178& 0.1942& 0.3339& 0.0659& 0.1649&0.0131\\
         8 & 17&16&  164.68&  6.95& 1.4777& 0.2040& 0.3027& 0.0656& 0.1367&0.0216\\
         9 & 28&32&  2878.51&  233.34& 1.3902& 0.2785& 0.3127& 0.0916& 0.1075&0.0270\\
         10 & 49&64&  96659.14&  7519.34& 1.3976& 0.2591& 0.2630& 0.0939& 0.0948&0.0200\\ \bottomrule
    \end{tabular}
    \caption{\textit{Mean simulation times and mean parameter values with standard deviations for NV-QWOA runs for problem sizes $5,6,7,8,9,10$}}
    \label{tab:sim-full-results}
\end{table} 

The average optimized parameter values for each problem size are also recorded in Table~\ref{tab:sim-full-results}. The large mean and standard deviation values for $\gamma$ in the $n = 6$ instances are attributed to a single instance identifying a region where $\gamma$ was an order of magnitude larger than the expected $\gamma \approx 1$, consequently leading to smaller values for $\beta$ and $t$. However, $\bar{\beta}$ and $\bar{t}$ decrease in proportion with $n$ as expected, and $\bar{\gamma}$ remains on the order of $\gamma \approx 1$, consistent with the findings in \cite{Bennett2024b}.
The average sum of objective function evaluations across instances (rounded up) was 645, 1199, 1902, 3134, 5030, and 8681 for each $n$, respectively; these values are linearly proportional to the growth of $p$ across problem sizes. As shown in Section~\ref{obj-func-eval-results}, the number of objective function evaluations reflects the polynomial scaling of $p$ with $n$.


\subsection{Number of objective function evaluations} \label{obj-func-eval-results}
\begin{figure}[ht]
    \centering
    \includegraphics[width=0.4\linewidth]{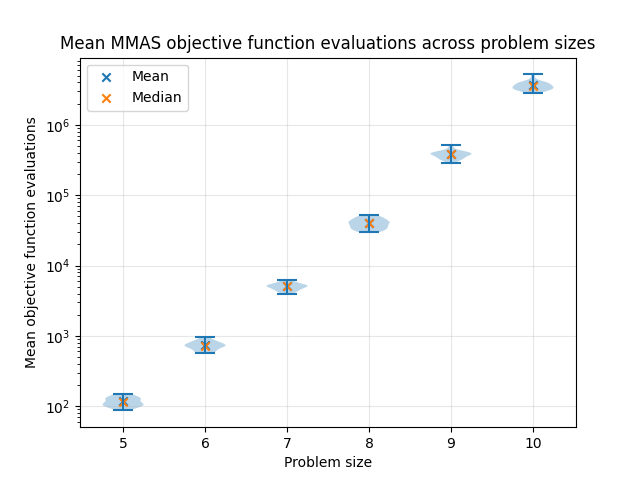}
    \caption{\textit{The mean number of objective function evaluations for the MMAS algorithm over problem sizes $5,6,7,8,9,10$}}
    \label{fig:mmas-obj-func-evals}
\end{figure}

The mean number of objective function evaluations, $N_{f(x)} = m \cdot I$, as a function of problem size $n$ for the MMAS is shown in Figure \ref{fig:mmas-obj-func-evals}. Both the average-case and worst-case number of evaluations clearly scale exponentially with the problem size, increasing by approximately one order of magnitude with each linear increase in $n$. This behavior is expected given the factorial growth of the solution space. As $n$ increases, the ratio of the fixed number of ants $m$ to the $n!$ possible solutions decays drastically; consequently, a significantly larger $I$ is required to compensate. Because the necessary increase in $I$ far outpaces the constant $m$, a sharp escalation in $N_{f(x)}$ is observed.

The mean number of objective function evaluations for the GLS is significantly higher than for the NV-QWOA across all problem sizes, as illustrated in Figure~\ref{fig:greedy-NV-QWOA-solve-measurement}. The GLS is designed to select the best improving swap among its neighbors; however, it does not consider the global cost landscape, nor does it terminate its comparison with neighboring solutions early. These characteristics lead to a high number of objective function evaluations. This could potentially be mitigated by augmentations to its search procedure, such as introducing randomisation to occasionally accept inferior solutions and avoid converging to local minima. 

However, such modifications introduce additional complexity through sophisticated mechanisms intended to escape local minima and enhance search quality. The implementation of these augmentations typically results in even more objective function evaluations, as observed in Breakout Local Search \cite{BENLIC20134800} and Tabu Search \cite{GLOVER1986533}. Consequently, while the standard GLS is more likely to terminate at a local minimum, it remains more efficient regarding the total number of evaluations compared to its more complex variants.

The difference in the number of evaluations between the GLS and NV-QWOA increases alongside the problem size. This indicates that the quantum heuristic exhibits a slower growth rate in evaluations, suggesting sub-exponential scaling across the smaller problem sizes investigated. In summary, the NV-QWOA achieves superior scaling in objective function evaluations compared to the MMAS and GLS heuristics.

\subsection{Solve and measurement probabilities}
\begin{figure}[ht]
    \centering
    \includegraphics[width=0.4\linewidth]{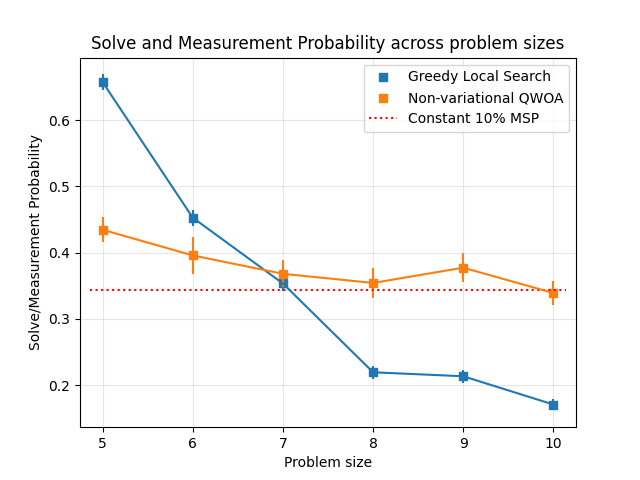}
    \includegraphics[width=0.4\linewidth]{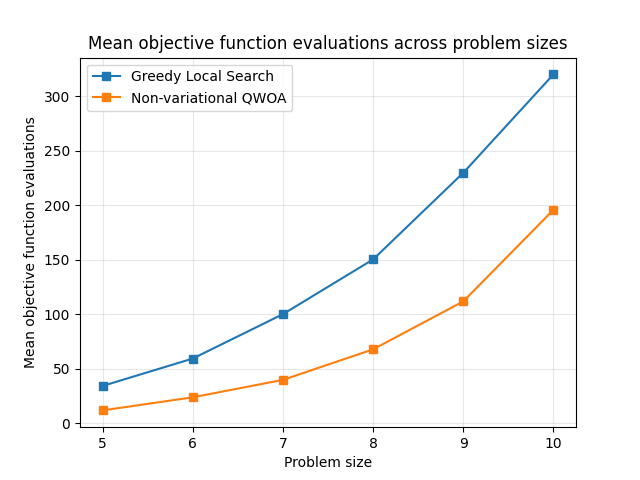}
    \caption{\textit{Direct comparison of GLS with NV-QWOA. Left: The solve and measurement probability for the GLS and NV-QWOA respectively over problem sizes $5,6,7,8,9,10$. Right: The mean number of objective function evaluations for the GLS and NV-QWOA over problem sizes $5,6,7,8,9,10$}}
    \label{fig:greedy-NV-QWOA-solve-measurement}
\end{figure}

The solve and measurement probabilities across problem sizes for the GLS and NV-QWOA are presented in Figure~\ref{fig:greedy-NV-QWOA-solve-measurement}. This figure includes a baseline representing the ideal case, where the NV-QWOA maintains a consistent 10\% Success Probability (OSP) for every instance when preparing and measuring the amplified state four times. The error bars indicate substantially greater variance in the NV-QWOA results compared to the GLS, suggesting that the GLS is more consistent in its performance across different instances of the same size.

Initially, for problem sizes $n \in \{5, 6\}$, the solve probability for the GLS is considerably higher than that of the NV-QWOA. However, as expected, the NV-QWOA consistently maintains its 10\% OSP across all problem sizes, whereas the solve probability for the GLS decays exponentially as $n$ increases. This decay is typical for local search heuristics; as the number of local minima grows exponentially with problem size, the probability of a greedy search converging to the global optimum decreases accordingly. Furthermore, the high initial solve probability for small $n$ can be attributed to the restricted solution space, which increases the likelihood that a random initial permutation is within a few swaps of the global minimum.

\subsection{Required iterations with problem size}
\begin{figure}[t]
    \centering
    \includegraphics[width=.45\linewidth]{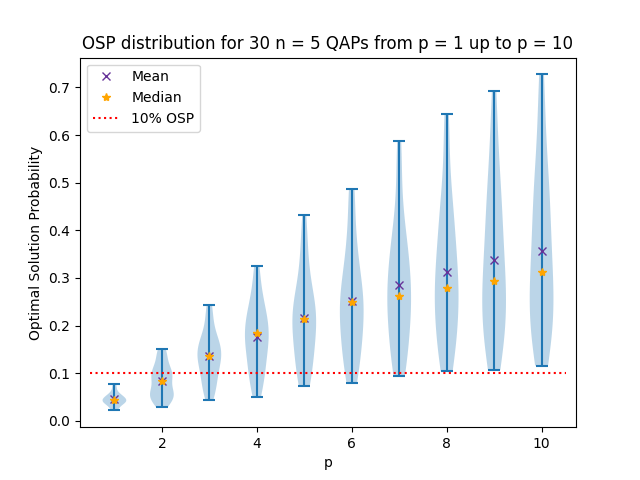}
    \includegraphics[width=.45\linewidth]{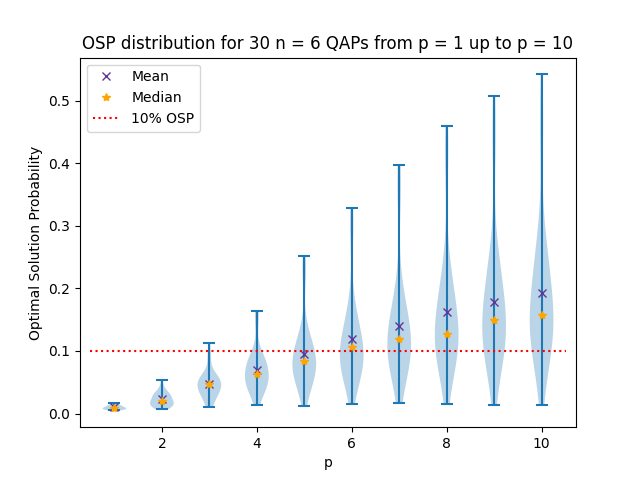}
    \includegraphics[width=.45\linewidth]{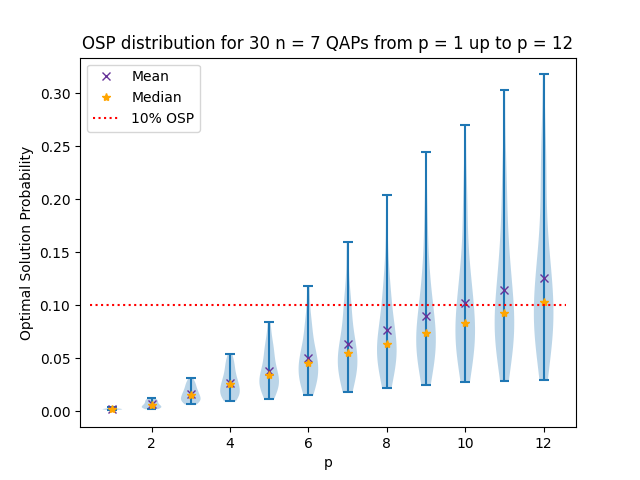}
    \includegraphics[width=.45\linewidth]{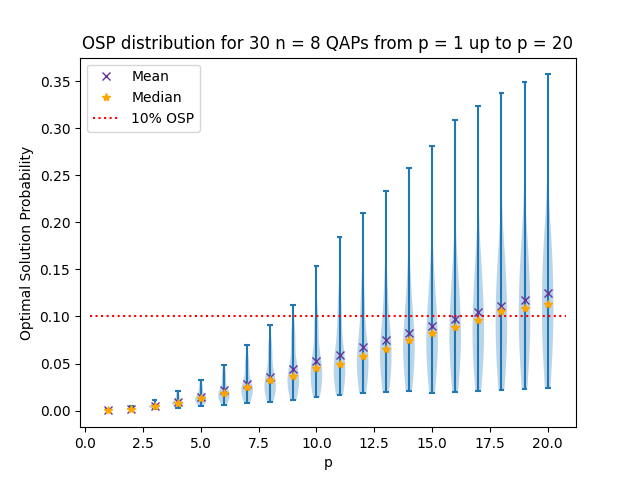}
    \includegraphics[width=.45\linewidth]{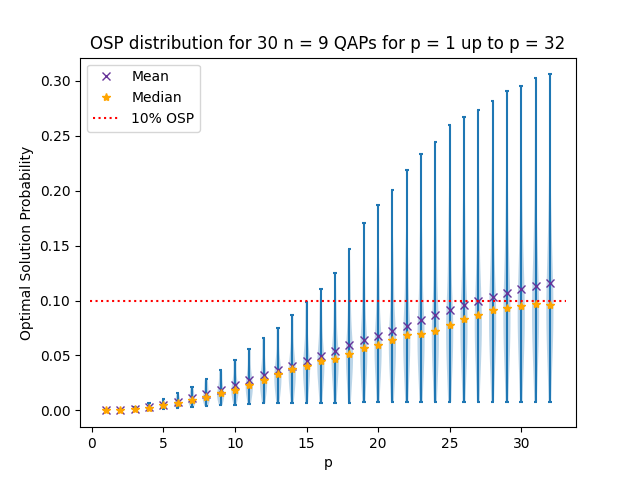}
	\includegraphics[width=.45\linewidth]{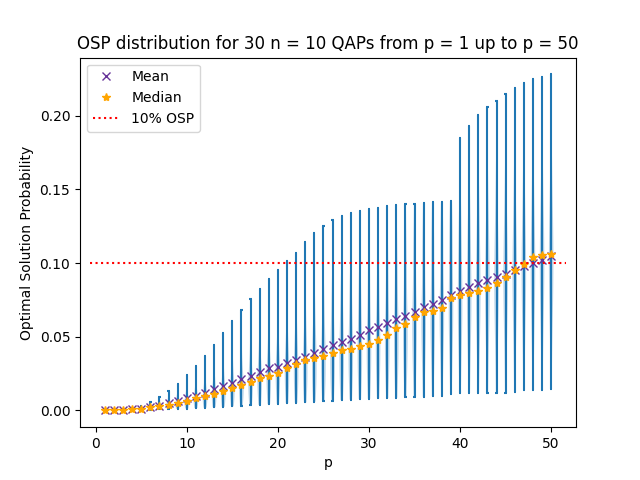}
	\caption[Optimal solution probability distribution of 30 random QAPs for different problem sizes]{\textit{Optimal solution probability violin plots across various $p$ value ranges}}
	\label{fig:Fig1}
\end{figure}

The distributions of OSPs across a varying scale for the depth $p$ of interleaving unitaries for the 30 random QAP instances generated for each problem size are shown as violin plots in Figure \ref{fig:Fig1}, along with the means and medians for the distributions at each discrete $p$ value. A 10\% OSP threshold line is also visible to see the optimal integer $p$ value that produces mean OSP $>$ 10\% in the average case for a given problem size. 

For $n=5$ up to $n=9$ instances, the distribution smoothly increases upwards to larger $p$ values as expected for the algorithm, with some flattening of the maximum OSPs observed for the larger problem sizes, most pronounced with the $n=9$ instances. For the $n=10$ instances, the flattening of maximum OSP is more pronounced around $p = 25$ before a sharp increase is observed at $p = 40$, with this maximum OSP continuing to increase with larger $p$. Observing the instances around this transition point, it is found that the maximum OSP is not consistent with a singular instance, but multiple. The parameter values for these instances significantly changed, indicating the optimiser found a nearby region of lower cost within the landscape. 

\begin{figure}[ht]
    \centering
    \includegraphics[width=0.4\linewidth]{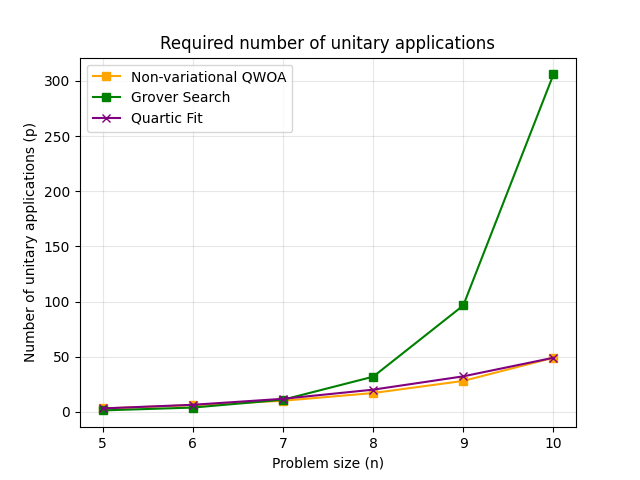}
    \caption{\textit{Required integer number of NV-QWOA iterations $p$ compared against Grover's search with problem sizes $n = 5, 6, 7, 8, 9, 10$}}
    \label{fig:unitary-applications-grover}
\end{figure}

The required number of alternating unitary applications for the NV-QWOA is then compared with the number of iterations of Grover's search to amplify OSP to 10\% in the average-case, as seen in Figure \ref{fig:unitary-applications-grover}. The two heuristics initially have similar iteration counts for low $n = 5, 6$, but after $n = 7$ is when the number of required rotations start to exponentially outnumber what is required for the NV-QWOA. Since the number of Grover iterations is governed by $p = \frac{\sqrt{N}\arcsin{(\sqrt{0.1})} - 1}{2}$, the number of required iterations to reach 10\% OSP is on order of $O(\sqrt{n!})$, and thus explains the exponential increase. We observe optimal integer iteration counts p = 3, 6, 10, 17, 28, 49 for the NV-QWOA, with quick curve fitting suggesting this optimal number of iteration counts scales quartically by $p = 0.0049 n^{4}$, which is of $O(n^4)$ with the problem size, in line with the hope of a polynomial scaling for the NV-QWOA for QAP instances.

\subsection{Solution shell probabilities and internode distance} \label{solution-shells-sec}
We will now consider the distribution of QAP solutions across the shells given a fixed vertex at the optimal solution across different problem instances following execution of the NV-QWOA. The solution shell probabilities is defined as the sum of probabilities of all solutions in a given shell at some distance i.e. the number of transpositions (permutation element swaps) away from the optimal solution. The internode distance is a solution's distance away from the optimal solution.  

\begin{figure}[ht]
    \centering
    \includegraphics[width=0.4\linewidth]{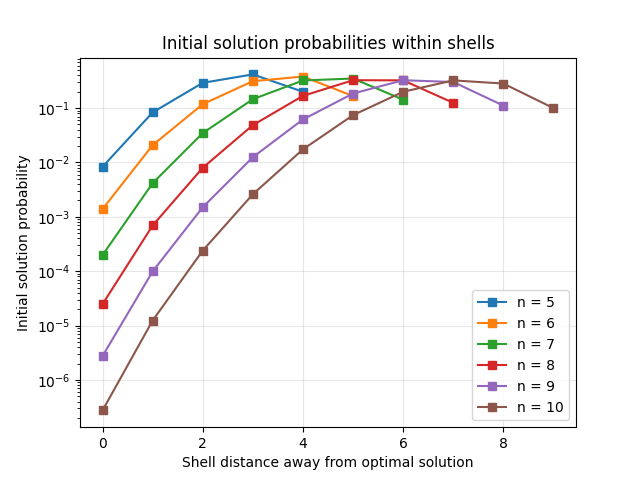}
    \includegraphics[width=0.4\linewidth]{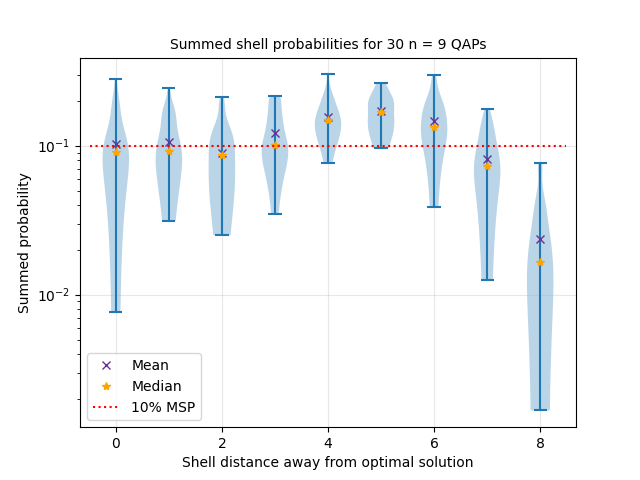}
    \caption{\textit{Left: The sum of initial probabilities considering shell distance away from optimal solution over problem sizes $5,6,7,8,9,10$. Right: The sum of final probabilities across the 30 $n = 9$ QAP instances}}
    \label{fig:initial-shell-probabilities}
\end{figure}

The initial probabilities of all solutions prior to execution of the interleaving unitary application are equal, given that the initial quantum state is an equal superposition of all possible solution states encoding solutions per Equation \ref{eq:initial-sup-eq}. From the number of solutions within each shell, considering distance away from the optimal solution, the sum of the initial probabilities for each shell is simply the number of solutions within that shell. The solutions in the $k^{th}$ shell are the permutations that have $n-k$ disjoint cycles. A plot of the sum of initial probabilities in each shell is given in Figure \ref{fig:initial-shell-probabilities}.

Clearly, the optimal solution probability (shell distance $= 0$) decays by $1/n$ in accordance with initial probability $1/n!$, and the NV-QWOA needs to provide drastic amplification to these earlier shells regardless of problem instance. For the $n = 9$ case, the amplification is $10^4$ fold, adequately boosting OSP to a more measurable regime in the average case. 

\begin{figure}[ht]
    \centering
    \includegraphics[width=0.4\linewidth]{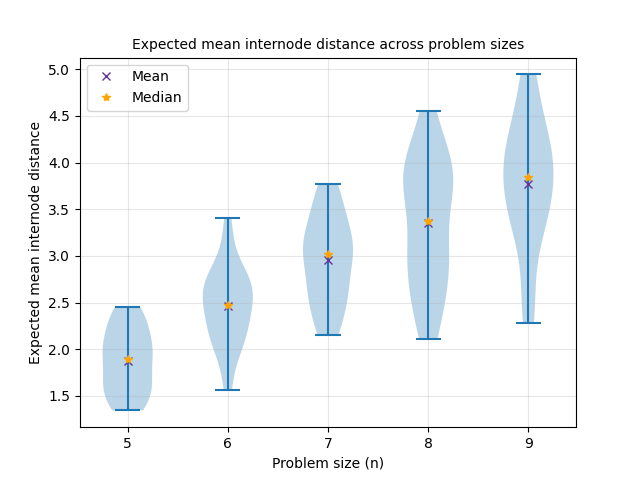}
    \caption{\textit{The expected mean of the internode distance across problem sizes $n = 5, 6, 7, 8, 9$}}
    \label{fig:exp-internode-distance}
\end{figure}

We expect that the NV-QWOA significantly amplifies the optimal and near-optimal solutions, whilst simultaneously attenuating probability of low-quality solutions across problem sizes, albeit with stronger effects needed for larger problem sizes. Figure \ref{fig:exp-internode-distance} shows the expected mean of the inter-node distance from the optimal solution for the 30 instances across the problem sizes of interest. This expected mean communicates the distance away from the optimal solution we expect the measured solution will be following execution of the NV-QWOA with the required $p$ circuit depth for a given problem size. The ideal expected mean would be $0$ and size-independent to guarantee finding the optimal solution with only one measurement with any problem size. However, the NV-QWOA is inherently a heuristic due to the complex phase interactions imposed from the randomised nature of amplitude probability transfer and phase-shifting across solution states from CTQWs. Optimal and near-optimal solutions will not consistently be amplified by the same proportion with every iteration of the NV-QWOA, and stronger amplification within factorial scaling is required to boost OSP to 10\% consistently. The expected means obtained show a linear scaling with problem size which, in comparison with the growth of number of shells with problem sizes, is promising. As the number of shells increase linearly with problem size, the expected mean internode distance initially is much greater for worst solutions.

\subsection{Discussion}

Our findings showcase that the NV-QWOA has competitive scaling against known efficient classical heuristics across 30 randomly generated QAP instances of small problem sizes with regards to number of objective function evaluations and measurement/solve probability. As seen in Figure \ref{fig:unitary-applications-grover}, the number of alternating unitary applications scales quadratically with problem size, and outperforms scaling against the number of Grover iterations required to amplify OSP to 10\% in the average-case, which scales with $O(\sqrt{n!})$. This is a promising result since $p$ grows linearly with circuit complexity of the NV-QWOA, and thus overall circuit complexity is polynomial in time over small QAP instances. Additionally, due to how Grover's search offers a quadratic speedup over exhaustive search, it suggests, at the very least, that the NV-QWOA might offer a greater-than-quadratic speedup in solving QAP instances of small sizes.

With the comparison of the quantum heuristics, only one metric was considered, and quantum noise was not embedded into the model. With regards to lack of noise modelling, simulations were focused on ideal quantum algorithm performance using fault-tolerant quantum devices due to the availability of optimisation software targeted for combinatorial optimisation compatible with parallel computation on Setonix. This focus on ideal performance is not aligned with the state of current NISQ-era hardware that involves algorithm design around noise mitigation, and thus such results are only indirectly impactful for real-world practical applications of quantum heuristics. The simulation results however serve as important theoretical benchmarks to later inform practical design when considering decoherence and noisy systems.  

There were several limitations with simulation of larger problem instances due to the intricacies of parallel computation in performing classical Hamiltonian simulations of quantum algorithms. The large simulation times introduced complications with efficiency in collecting these results, and it is expected the rapid increase in simulation time will persist for larger problem sizes, requiring in-depth profiling of code, stronger parallel exploitation of the mixing graph, and even GPU-accelerated runs to investigate bottlenecks in computation, and allow to scale accordingly with problem size.  

Due to these small problem sizes, there are questions of whether the results obtained for the NV-QWOA would hold true for larger problem sizes, especially in the $n > 30$ regime where exact algorithms become too computationally intensive and thus intractable for practical use, soliciting heuristic use. In related studies with optimising parameter initialisation for the QAOA, several researchers found that the parameters of smaller problem instances could be extrapolated for use on larger problems \cite{galda2021transferabilityoptimalqaoaparameters}, substantially reducing computation time for such sizes, and inviting insight into heuristic performance within this regime.

\section{Conclusion and future work}

This paper explored the performance of the NV-QWOA in solving the QAP, with comparison of its behaviour conducted with two classical heuristics, the Max-Min Ant System (MMAS) and Greedy Local Search (GLS), as well as a brief comparison with Grover's search algorithm with the number of algorithm iterations needed to amplify the optimal solution probability (OSP) to 10\% for near-guaranteed measurement over multiple shots.     
30 QAP instances with sizes $n = 5, 6, 7, 8, 9, 10$ were randomly generated to define the instance set to be solved with the heuristics. The performance of the NV-QWOA over this instance set shows polynomial scaling of circuit depth with small problem sizes, and offers competitive performance when compared to the heuristics. 

Three metrics were used to qualify and quantify the performance of heuristics under different pairwise comparisons: the number of objective function evaluations between MMAS and GLS against the NV-QWOA, the measurement/solve probability of the GLS against the NV-QWOA, and the required number of algorithm iterations of Grover search against the NV-QWOA.

For future work, it is useful to investigate whether these trends persist for larger QAP sizes beyond $n = 30$ to see whether quantum heuristics offer an advantage in solving NP-hard problems more efficiently. There lies hardware concerns with the enormous parallel computation overhead for such problem sizes, given the computation time needed for the smaller instances explored here, and how the solution space grows factorially with $n$. However, it is expected the polynomial scaling results will persist for larger sizes, since the interference effect of the NV-QWOA relies on statistical assumptions that are increasingly valid for such sizes \cite{Bennett2024b}, and there is growing interest in developing reliable parameter transfer schemes to utilise values obtained from small instances for larger instances in the classically intractable regime. 

\section*{Acknowledgments}

This project was supported by the Australian Government via the Critical Technologies Challenge Program (CTCP) and the Advanced Strategic Capabilities Accelerator (ASCA). Substantial computational resource was provided by the Pawsey Supercomputing Research Centre and Acacia Object Storage. The authors would like to thank Tavis Bennett for valuable discussions on the quantum optimisation methodologies, and Edric Matwiejew for assistance with the implementation of the simulation code through use of the QuOp\_MPI package \cite{Matwiejew_2022, quopmpisite}, and fruitful discussions about quantum combinatorial optimisation. 

\newpage
\bibliographystyle{unsrt}
\bibliography{jbw1, References}

@article{christiansen2025instancespaceanalysisquadratic,
  title={Instance Space Analysis for the Quadratic Assignment Problem},
  author={J. Christiansen \& K. Smith-Miles},
  journal={\href{https://arxiv.org/abs/2506.20172}{arXiv:2506.20172}},
  year={2025}
}

@article{kooopmans1957,
title={Assignment Problems and the Location of Economic Activities},
author={T. C. Koopmans \& M. Beckmann},
journal={Econometrica},
volume={25},
number={1},
year={1957},
pages={53-76}}

@article{bazaraa1980,
author = {M. S. Bazaraa \& H. D. Sherali},
title = {Benders' partitioning scheme applied to a new formulation of the quadratic assignment problem},
journal = {Naval Research Logistics Quarterly},
volume = {27},
number = {1},
pages = {29-41},
year = {1980}
}

@article{bittel2021,
  title = {Training Variational Quantum Algorithms Is NP-Hard},
  author = {L. Bittel \& M. Kliesch},
  journal = {Phys. Rev. Lett.},
  volume = {127},
  pages = {120502},
  numpages = {6},
  year = {2021}
}

@article{loiola2007,
    author = {E. M. Loiola et. al}, 
    title = {A survey for the quadratic assignment problem},
    journal = {European Journal of Operational Research},
    volume = {176},
    pages = {657-690},
    year = {2007}
}

@article{gilmore1962,
  title={Optimal and Suboptimal Algorithms for the Quadratic Assignment Problem},
  author={P. C. Gilmore},
  journal={Journal of The Society for Industrial and Applied Mathematics},
  year={1962},
  volume={10},
  pages={305-313}
}

@article{lawler1963,
author = {E. L. Lawler},
title = {The Quadratic Assignment Problem},
journal = {Management Science},
volume = {9},
number = {4},
pages = {586-599},
year = {1963}
}

@article{sahni1976,
    author = {S. Sahni \& T. Gonzalez},
    title = {P-Complete Approximation Problems},
    journal = {Journal of the Association for Computing Machinery},
    volume = {23},
    number = {3},
    pages = {555-565},
    year = {1976}
}

@article{elshafei1977,
    author = {A. N. Elshafei},
    title = {Hospital Layout as a Quadratic Assignment Problem},
    journal = {Opl Res. Q.},
    volume = {28},
    number = {1},
    pages = {167-179},
    year = {1977}
}

@article{steinberg1961,
    author = {L. Steinberg},
    title = {The Backboard Wiring Problem: A Placement Algorithm},
    journal = {SIAM Review},
    volume = {3},
    number = {1},
    pages = {37-50},
    year = {1961}
}

@article{geoffrion1976,
    author = {A. M. Geoffrion \& G. W. Graves},
    title = {Scheduling Parallel Production Lines with Changeover Costs: Practical Application of a Quadratic Assignment / LP approach},
    journal = {Operations Research},
    volume = {24},
    number = {4},
    pages = {595-610},
    year = {1976}
}

@article{anstreicher2002,
    author = {K. M. Anstreicher et. al},
    title = {Solving large quadratic assignment problems on computational grids},
    journal = {Math. Program. Ser. B},
    volume = {91},
    pages = {563-588},
    year = {2002}
}

@article{farhi2000,
    author = {E. Farhi \& J. Goldstone \& S. Gutmann \& M. Sipser},
    title = {Quantum computation by Adiabatic Evolution},
    journal = {\href{https://arxiv.org/abs/quant-ph/0001106}{ arXiv:quant-ph/0001106}},
    year = {2000}
}

@article{farhi2001,
    author = {E. Farhi et. al},
    title = {A Quantum Adiabatic Evolution Algorithm Applied to Random Instances of an NP-Complete Problem},
    journal = {Science},
    volume = {292},
    number = {5516},
    pages = {472-475},
    year = {2001}
}

@misc{quopmpisite,
    author = {E. Matwiejew},
    title = {QuOp\_MPI v.1.2.1 - https://github.com/Edric-Matwiejew/QuOp\_MPI},
    year = {2025},
    url = {https://arxiv.org/abs/2404.03167}
}

@misc{qaplib,
    key = {QAPLIB 2012},
    title = {\href{https://coral.ise.lehigh.edu/data-sets/qaplib/qaplib-problem-instances-and-solutions/}{QAPLIB 2012}}
}

@article{gavett1966,
    author = {J. W. Gavett \& N. V. Plyter},
    title = {The Optimal Assignment of Facilities to Locations by Branch and Bound},
    journal = {Operations Research},
    volume = {14},
    number = {2},
    pages = {210-232},
    year = {1966}}

@article{padberg1991,
    author = {M. W. Padberg \& G. Rinaldi},
    title = {A branch-and-cut algorithm for the resolution of large-scale symmetric traveling salesman problems},
    journal = {SIAM Review},
    volume = {33},
    pages = {60-100},
    year = {1991}
}

@article{peng2010,
    author = {J. Peng \& H. Mittelmann \& X. Li},
    title = {A New Relaxation Framework for Quadratic
Assignment Problems based on Matrix Splitting},
    journal = {Math. Prog. Comp.},
    volume = {2},
    pages = {59-77},
    year = {2010}
}

@article{merz2002,
    author = {P. Merz \& B. Freisleben},
    title = {Greedy and Local Search Heuristics for Unconstrained Binary Quadratic Programming},
    journal = {Journal of Heuristics},
    volume = {8},
    pages = {197-213},
    year = {2002}
}

@inbook{ils,
    author = {H. R. Lourenço \& O. Martin \& T. Stützle},
    title = {Iterated Local Search: Framework and Applications},
    bookTitle = {Handbook of Metaheuristics},
    publisher = {Kluwer Academic Publishers},
    journal = {International Series in Operations Research \& Management Science},
    volume = {146},
    pages = {363-397},
    year = {2010},
}

@article{kirkpatrick1983,
    author = {S. Kirkpatrick \& C. D. Gelatt \& M. P. Vecchi},
    title = {Optimization by Simulated Annealing},
    journal = {Science},
    volume = {220},
    number = {4598},
    pages = {671-681},
    year = {1983}
}

@article{GLOVER1986533,
title = {Future paths for integer programming and links to artificial intelligence},
journal = {Computers \& Operations Research},
volume = {13},
number = {5},
pages = {533-549},
year = {1986},
author = {F. Glover}
}

@phdthesis{dorigo1992,
    author = {M. Dorigo},
    title = {Optimization, Learning and Natural Algorithms},
    school = {Politecnico di Milano, Italy},
    year = {1992}
}

@article{moscato2000,
author = {P. Moscato},
year = {2000},
pages = {},
title = {On Evolution, Search, Optimization, Genetic Algorithms and Martial Arts - Towards Memetic Algorithms},
journal = {Caltech Concurrent Computation Program}
}

@article{bartzbeielstein2014,
author = {T. Bartz-Beielstein \& J. Branke \& J. Mehnen \& O. Mersmann},
year = {2014},
pages = {},
title = {Evolutionary Algorithms},
volume = {4},
journal = {Wiley Interdisciplinary Reviews: Data Mining and Knowledge Discovery}
}

@article{STUTZLE2000889,
title = {MAX–MIN Ant System},
journal = {Future Generation Computer Systems},
volume = {16},
number = {8},
pages = {889-914},
year = {2000},
author = {T. Stützle \& H. H. Hoos}
}

@techreport{maniezzo1994,
    author = {V. Maniezzo \& M. Dorigo \& A. Colorni},
    title =  {The ant system
applied to the quadratic assignment problem},
    institution = {Université de Bruxelles, Belgium},
    year = {1994}
}

@article{Hadfield_2019,
   title={From the Quantum Approximate Optimization Algorithm to a Quantum Alternating Operator Ansatz},
   volume={12},
   number={2},
   journal={Algorithms},
   publisher={MDPI AG},
   author={S. Hadfield et. al},
   year={2019},
   pages={34}}

@misc{scipy,
    key = {SciPy},
    author = {E. Jones \& T. Oliphant \& P. Peterson},
    title = {\href{http://www.scipy.org/}{SciPy: Open source scientific tools for
    Python}},
    year = {2001}
}

@article{Matwiejew_2022,
   title={QuOp\_MPI: A framework for parallel simulation of quantum variational algorithms},
   volume={62},
   journal={Journal of Computational Science},
   publisher={Elsevier BV},
   author={E. Matwiejew \& J. B. Wang},
   year={2022},
   pages={101711}
}

@article{BENLIC20134800,
title = {Breakout local search for the quadratic assignment problem},
journal = {Applied Mathematics and Computation},
volume = {219},
number = {9},
pages = {4800-4815},
year = {2013},
author = {U. Benlic \& J. Hao}
}

@article{galda2021transferabilityoptimalqaoaparameters,
      title={Transferability of optimal QAOA parameters between random graphs}, 
      author={A. Galda et. al},
      journal={\href{https://arxiv.org/abs/2106.07531}{arXiv:2106.07531}},
      year={2021}
}

@conference{slurm-cite,
    author = {M. A. Jette \& T. Wickberg},
    booktitle = {Klusáček, D., Corbalán, J., Rodrigo, G.P. (eds) Job Scheduling Strategies
for Parallel Processing. JSSPP 2023. Lecture Notes in Computer Science},
    title = {Architecture of the Slurm Workload Manager},
    publisher = {Springer, Cham},
    volume = {14283},
    year = {2023}
}

@article{taillard1991,
    author = {E. D. Taillard},
    title = {Robust taboo search for the quadratic assignment problem},
    journal = {Parallel Computing},
    volume = {17},
    pages = {443-455},
    year = {1991}}

@article{taillard1995,
    author = {E. D. Taillard},
    title = {Comparison of iterative searches for the Quadratic Assignment Problem},
    journal = {Location Science},
    volume = {3},
    pages = {87-105},
    year = {1995}}

@inproceedings{Bennett2024a,
  title={Non-variational Quantum Combinatorial Optimisation},
  author={T. Bennett \& L. Noakes \& J. B. Wang},
  booktitle={\href{https://arxiv.org/abs/2404.03167}{arXiv:2404.03167} and 2024 IEEE International Conference on Quantum Computing and Engineering (QCE)},
  volume={1},
  pages={31--41},
  year={2024},
  organization={IEEE}
}

@article{Bennett2024b,
  title={Analysis of the non-variational quantum walk-based optimisation algorithm},
  author={T. Bennett \& L. Noakes \& J. B. Wang},
  journal={\href{https://arxiv.org/abs/2408.06368}{arXiv:2408.06368}},
  year={2024}
}

@article{marsh2021,
  title={Deterministic spatial search using alternating quantum walks},
  author={S. Marsh \& J. B. Wang},
  journal={Physical Review A},
  volume={104},
   pages={022216},
  year={2021},
  publisher={APS}
}

@article{Marsh2019,
  year = {2019},
  publisher = {Springer Science and Business Media {LLC}},
  volume = {18},
  number = {3},
  pages = {61},
  author = {S. Marsh \& J. B. Wang},
  title = {A quantum walk-assisted approximate algorithm for bounded {NP} optimisation problems},
  journal = {Quantum Information Processing}
}

@article{matwiejew_quantum_2023,
	title = {Quantum optimisation for continuous multivariable functions by a structured search},
	volume = {8},
	number = {4},
	journal = {Quantum Science and Technology},
	author = {E. Matwiejew \& J. Pye \& J. B. Wang},
	year = {2023},
	pages = {045013},
}

@article{Grover1997,
  title = {Quantum Mechanics Helps in Searching for a Needle in a Haystack},
  volume = {79},
  number = {2},
  journal = {Physical Review Letters},
  publisher = {American Physical Society (APS)},
  author = {L. K. Grover},
  year = {1997},
  pages = {325–328}
}

@article{slate2021quantum,
  title={Quantum walk-based portfolio optimisation},
  author={N. Slate \& E. Matwiejew \& S. Marsh \& J. B. Wang},
  journal={Quantum},
  volume={5},
  pages={513},
  year={2021},
  publisher={Verein zur F{\"o}rderung des Open Access Publizierens in den Quantenwissenschaften}
}

@article{farhi2014quantumapproximateoptimizationalgorithm,
      title={A Quantum Approximate Optimization Algorithm}, 
      author={E. Farhi \& J. Goldstone \& S. Gutmann},
      journal={\href{https://arxiv.org/abs/1411.4028}{arXiv:1411.4028}},
      year={2014}
}

@article{Marsh2020,
   title={Combinatorial optimization via highly efficient quantum walks},
   volume={2},
   number={2},
   journal={Physical Review Research},
   publisher={American Physical Society (APS)},
   author={S. Marsh \& J. B. Wang},
   year={2020}}

@inproceedings{grover1996,
  title={A fast quantum mechanical algorithm for database search},
  author={L. K. Grover},
  booktitle={Proceedings of the 28th Annual ACM Symposium on Theory of Computing (STOC)},
  pages={212--219},
  year={1996},
  organization={ACM}
}

@article{Larocca2024ReviewBP,
  title = {A Review of Barren Plateaus in Variational Quantum Computing},
  author = {M. Larocca et. al},
  journal = {\href{https://doi.org/10.48550/arXiv.2405.00781}{ arXiv:2405.00781}},
  year = {2024}
}

@article{Qu2024,
  title={Experimental implementation of quantum-walk-based portfolio optimization},
  author={D. Qu et. al},
  journal={Quantum Sci. Technol.},
  volume={9},
   pages={025014},
  year={2024}}

@article{Bennett2025,
  title={Benchmarking Quantum Heuristics:
Non-Variational QWOA for Weighted Maxcut},
  author={T. Bennett \& A. Smith \& E. Matwiejew \& J. B Wang},
  journal={Physical Review A (in press)},
  volume={ },
  pages={},
  year={2026},
}

@article{Bennett2021quantum,
  title={Quantum walk-based vehicle routing optimisation},
  author={T. Bennett \& E. Matwiejew \& S. Marsh \& J. B. Wang},
  journal={Frontiers in Physics},
  volume={9},
  pages={730856},
  year={2021},
  publisher={Frontiers Media SA}
}

@book{QWbook2014,
	title        = {Physical implementation of quantum walks},
	author       = {K. Manouchehri \& J. B. Wang},
	year         = 2014,
	publisher    = {Springer},
	options      = {extsym={*}}
}

@article{LokeWang2017,
  title={Efficient quantum circuits for continuous-time quantum walks on composite graphs},
  author={T. Loke \& J. B. Wang},
  journal={Journal of Physics A: Mathematical and Theoretical},
  volume={50},
  number={5},
  pages={055303},
  year={2017},
}


\end{document}